\documentclass{optica-article}

\journal{opticajournal} 

\articletype{Research Article}

\usepackage{lineno}
\usepackage{graphicx}
\usepackage{subcaption}
\usepackage{float}
\usepackage{placeins}
\usepackage{hyperref}

\begin{document}

\title{Guidance of ultraviolet light down to 190 nm in a hollow-core optical fibre}

\author{Robbie Mears,\authormark{1,*} Kerrianne Harrington,\authormark{1} William J. Wadsworth,\authormark{1} Jonathan C Knight,\authormark{1} James M Stone,\authormark{1} Tim A Birks,\authormark{1}}

\address{\authormark{1}Centre for Photonics and Photonic Materials, Department of Physics, University of Bath, Bath, BA2 7AY, UK\\}

\email{\authormark{*}rm2033@bath.ac.uk} 


\begin{abstract}
We report an anti-resonant hollow core fibre with ultraviolet transmission down to 190 nm, covering the entire UV-A, UV-B and much of the UV-C band. Guidance from 190 - 400 nm is achieved apart for a narrow high loss resonance band at 245 – 265 nm. The minimum attenuation is 0.13 dB/m at 235 nm and 0.16 dB/m at 325 nm. With an inscribed core diameter of \(\sim\)12 µm, the fibre's bend loss at 325 nm was 0.22 dB per turn for a bend radius of 3 cm at 325 nm.
\end{abstract}

\section{Introduction}
The flexible and robust transmission of ultraviolet (UV) light through optical fibres would be invaluable in reimagining a multitude of applications including ultrafast spectroscopy, precision laser machining and UV generation \cite{Kotsina2019,wolynski2011}. However, fibres with solid silica cores suffer increasing attenuation towards shorter wavelengths due to electronic absorptions and Rayleigh scattering. Furthermore, light-induced damage to the silica structure by energetic UV photons causes solarization \cite{YuUV}, which is particularly limiting for fibre applications in the far UV. Even specially designed solarization resistant fibres suffer from severe transmission degradation in use, experiencing 6 dB of induced loss after 10 minutes of exposure to 35 mW of incident power at 266 nm \cite{YuUV}.

Anti-resonant hollow core fibres (AR-HCFs) can circumvent these problems because the overlap of light with the glass is very small, with air having a low attenuation for much of the UV range. Only towards 200 nm do electronic absorptions by oxygen and water vapour \cite{waterabs1997,waterabs2,oxabs1953} rapidly increase the UV losses, setting the lower limit in air to \(\sim\)190 nm. The reduced overlap, and so absorption loss, has therefore extended the working range of silica-based fibres from the vacuum UV \cite{Winter19} to the mid-IR \cite{YuIR}, until material absorption is so great that even the much-reduced overlap of the light with glass causes significant attenuation. While this limit has been thoroughly investigated in the mid-IR \cite{YuIR}, the far UV remains largely unexplored. Efficient \textit{generation} of deep UV light has been reported in gas filled hollow core fibres, enabled by low losses at pump wavelengths \cite{Belli15,Ermalov1}. However, the fibres used are very short ( 15 – 25 cm ) and the UV light is only generated in the last few centimetres. There was no measurement of transmission losses, and no evidence that the fibres could transmit deep UV light over lengths of a metre or more.

In a simple AR-HCF, Fig. \ref{fig:fibres}, the thin walls of the silica capillaries surrounding the air core act as Fabry-Perot resonators, resulting in leaky confinement of light within "anti-resonant" wavelength ranges. High-loss resonance wavelengths \(\lambda\) are defined by \cite{YuIR}

\begin{equation}
     \lambda_{m} = \frac{2t}{m}(n^{2}-1)^\frac{1}{2} \label{eqn1}   
\end{equation}	 	
where resonance order \textit{m }is an integer, \textit{t} is the wall thickness and \textit{n} is the refractive index of silica. The broad low-loss anti-resonant bands lie in between. Higher order anti-resonant bands have reduced bandwidth and increased overlap of light and glass \cite{Fokoua23}, so the widest bandwidth occurs in the fundamental band (around \textit{m} = 0.5). 

While wall thickness is thus constrained, core size can be varied more widely. Larger cores reduce confinement loss (in the straight fibre) but increase bend loss. For real-world applications a balance can be found when the core diameter is 20 -- 30 times the wavelength \cite{YuHCF}. 

These parameters are difficult to achieve for the UV as fibre microstructures must be drawn smaller than those guiding in the visible or infrared. This greater draw-down ratio leads to increased deformation \cite{Chen13} due to surface tension as the fibre is drawn. While a fibre with a core wall thicknesses of 132 nm \cite{YuUV} has been reported, the fabrication challenge is shown by its low yield. Another fibre with a 110 nm \cite{Ding2019} wall showed an imperfect structure (two touching resonators forming a node \cite{Fokoua23}) with no measured transmission reported below 270 nm. 

One solution to the problems of high draw-down ratios is to separate the drawing process into several smaller steps or stages. In a typical stack-and-draw method \cite{YuIR}, there is a single intermediate preform stage between the stack and the fibre. Recent work has demonstrated the benefits of two intermediate stages \cite{Davidson23}, reducing the draw-down ratio at the final fibre stage and yielding greater control of the microstructure. Using this method, a fibre with high structural uniformity and 167 nm wall thickness has been demonstrated, with broadband guidance in the ultraviolet-visible range \(\sim\) 360 -- 1100 nm in the fundamental band. However, guidance further into the UV was not reported.

The penalty of this method is a more-awkward fabrication process with a smaller potential yield, since the final preform contains less glass. The extreme case of this is fibre tapering, which can create the desired structure through effectively many intermediate stages \cite{Birks1992,Pennetta2019}. While suitable for some vacuum-UV applications \cite{Winter19}, the yield is limited to only tens of centimetres at most.

An alternative approach is to increase the wall thickness (to use a higher-order anti-resonance band) and/or core size to reduce the required draw-down ratio. This has the effect of reducing the available bandwidth in the UV and increasing bend loss respectively, limiting the range of applications. Recent examples of UV guiding fibres \cite{osorio2023,Gao18,Leroi2023} had wall thicknesses of \(\sim\) 600 nm and core sizes of 15 – 27 µm, with no measured guidance below 250 nm. The 27 µm core reported in \cite{Leroi2023} displayed rapidly increasing bend loss for bend radii below 7 cm.

In summary, robust guidance at the shortest wavelengths in AR-HCFs has not previously been demonstrated. Most of the relevant literature does not report guidance below 270 nm at all \cite{Ding2019,Davidson23,osorio2023,Gao18,Leroi2023,Frosz23}. Furthermore, those fibres with large cores suffer from high bend loss \cite{Leroi2023,Frosz23} or are expected to do so \cite{osorio2023}; those with thicker walls \cite{Leroi2023,osorio2023,Gao18} guide only in narrow high-order bands; the complex fabrication procedure of \cite{Davidson23} limits yield; and the irregular capillary wall thickness of \cite{Ding2019} will broaden the high-loss resonance bands. Uniquely for drawn fibres, \cite{YuUV} does demonstrate guidance below 270 nm, but with key limitations for practical applications: a large core that is likely to cause high bend loss (though bend loss was not measured); a broad high-loss band across 230 - 300 nm (caused by the lower resonance order and perhaps structural asymmetry); and a short length/low yield. The tapered fibre in \cite{Winter19} is far too short for most applications.

Here we report an anti-resonant hollow-core fibre which covers much of the ultraviolet range possible in air, demonstrating single-mode guidance from 190 - 400 nm except for a narrow resonance at 245 – 265 nm. The attenuation is 0.13 dB/m at 235 nm and 0.16 dB/m at 325 nm. Bend loss measurements show 95\% of straight-fibre transmission at 325 nm for a single turn at a 3 cm bend radius. The fibre fabrication was made possible through improvements in fibre drawing, allowing a single intermediate preform stage, yielding a continuous length of 250 m; in principle, the 1 m preform contained enough glass to draw 5 km of fibre.
\section{Fabrication}

The fibre was fabricated using the stack-and-draw method with a single intermediate stage, with all elements made from Heraeus F300 low-OH synthetic silica which contains a relatively high proportion of chlorine. Despite this, we did not observe the presence of ammonium chloride detailed in \cite{Rikimi20} at any stage. Capillaries of 2034 ± 2 µm outer diameter were drawn from a tube of 21/25 mm (inner and outer diameters) and stacked in an outer tube of 11/18.7 mm with supporting tubes. The stack was then drawn to canes of 2720 ± 2 µm, where the resonators were sealed to induce a passive build-up of pressure during the draw. This prevents the uncontrolled collapse of resonators, improving their uniformity in the subsequent fibre drawing stage.

These canes were then placed in a 3/10 mm jacket and drawn to 140 µm diameter fibre, with a vacuum applied between cane and jacket to remove the interstitial region. The parameters for the reported fibre draw were a set furnace temperature of 1920\textdegree C, a feed rate of 18 mm/min and a draw rate of\(\sim\) 92 m/min, yielding a tension of 430 ± 15 g and a stress of 281 ± 10 MPa. The high drawing stress exceeds the 250 MPa considered high in \cite{Davidson23} while allowing an entire preform to be drawn without breaks. We find that stresses even twice as great as this are made possible without frequent fibre breaks through the use of higher feed and drawing speeds. In contrast, drawing similar fibres at slower speeds results in frequent fibre breaks for the same drawing tension. Given that an increased process speed entails a hotter furnace temperature for a given drawing tension, we expect that reduced exposure of the fibre to the lower temperature range where devitrification can occur (as it is drawn through the margin of the furnace hot-zone) may explain its enhanced strength \cite{devit}.

\begin{figure}[h]
\centering
\includegraphics[width=1\linewidth]{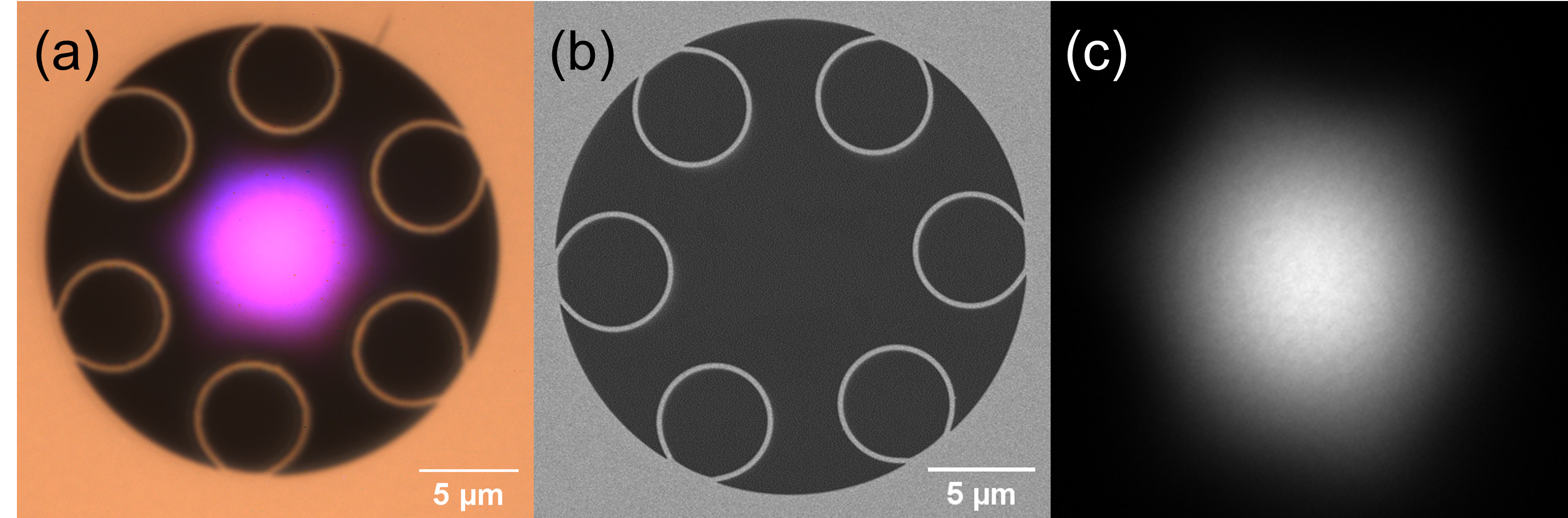}
\caption{(a) Optical micrograph of the fibre illuminated by transmitted light. (b) SEM image of the  fibre (c) Near-field image at the output of 17.6 m of the fibre through a 10 nm bandpass filter centred at 350 nm}
\label{fig:fibres}
\end{figure}

While at the given drawing conditions, a range of differential pressures between the cladding and core were applied to optimise the microstructure. Lengths from 100 - 1000 m were collected, with 250 m of the reported fibre was collected in a single length at a differential pressure of 12.65 kPa. Fig. \ref{fig:fibres}(a) is an optical micrograph, and Fig.\ref{fig:fibres}(b) is a scanning electron micrograph (SEM) of the fibre cross section.
\FloatBarrier
\section{UV guidance}
For all measurements an Energetiq EQ-99X laser-driven light source was used, with offset parabolic mirrors to couple the light into the fibre. Spectra were measured using a Bentham DTMc300 double monochromator and large-area UV photodetector. Near-field images were taken using a silicon camera appeared to show the fundamental mode with no apparent change on bending the fibre, Fig. \ref{fig:fibres}(c), suggesting effectively single-mode guidance.

\begin{figure}[h]
\centering
\includegraphics[width=1\linewidth]{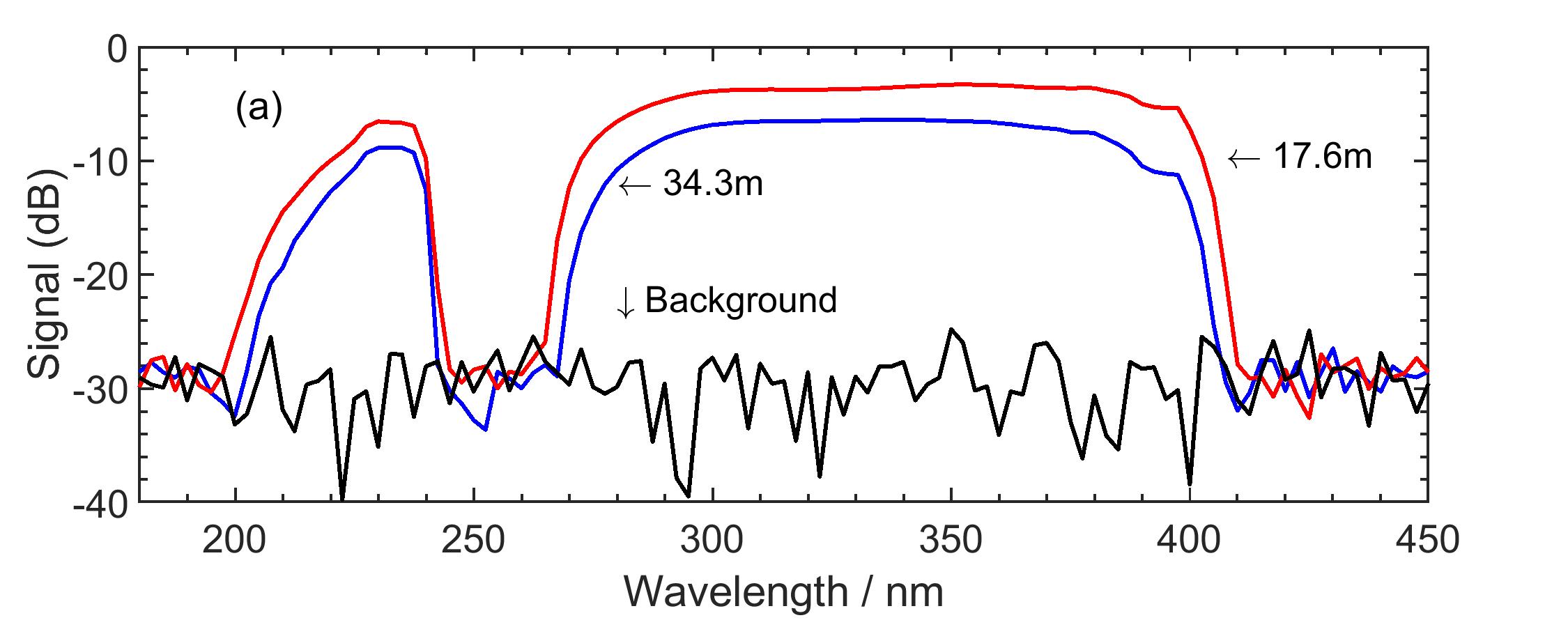}
\centering
\centering
\includegraphics[width=1\linewidth]{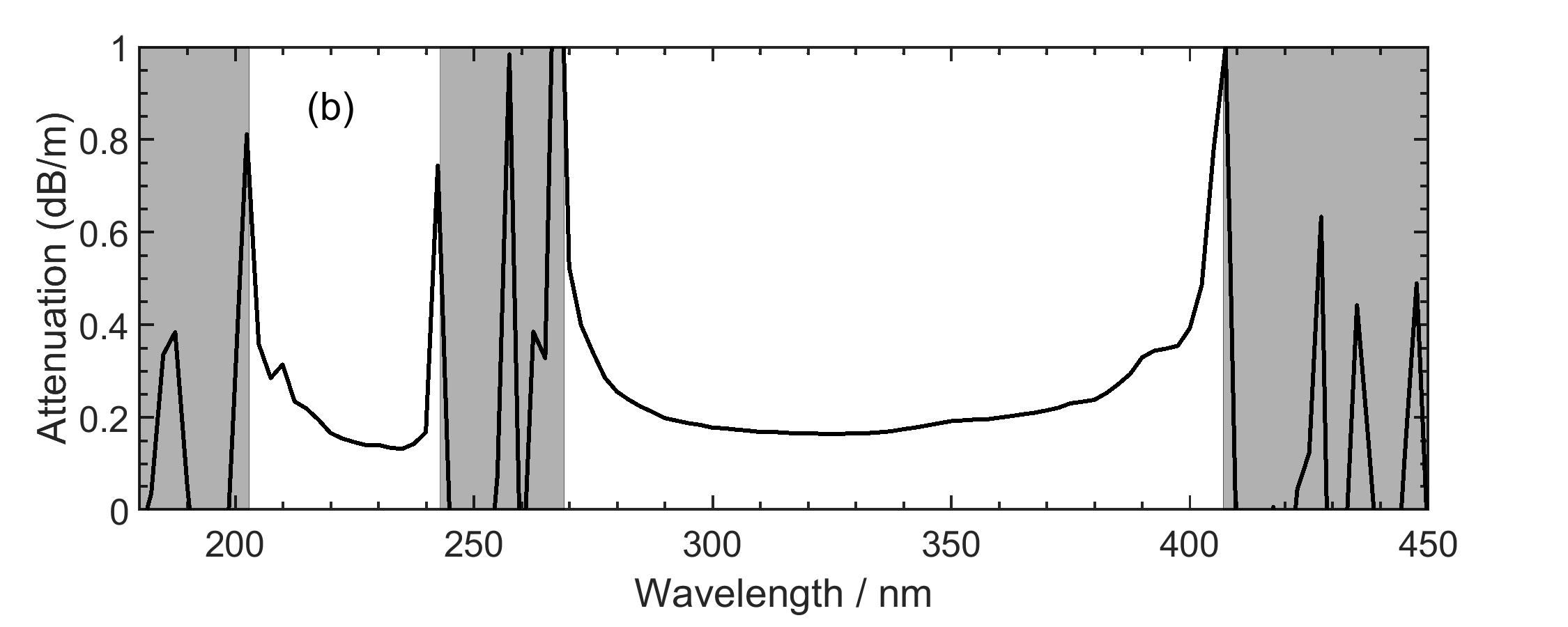}
\caption{(a) Transmission spectra (measured photodiode current) versus wavelength on a logarithmic scale for fibre lengths of 34.3 m and 17.6 m, with a background scan obtained by blocking the light. (b) Attenuation calculated from the transmission spectra. Shaded regions indicate no signal above noise}
\label{fig:longcutback}
\end{figure}

The fibre attenuation was measured using a cutback experiment. 34.3 m of fibre was wound in loose loops of \(\sim\)45 cm diameter on a table to minimise bend loss. Transmission spectra through the fibre, recorded before and after cutting the fibre back to 17.4 m, are shown in Fig. \ref{fig:longcutback} along with the attenuation. 

\FloatBarrier
The spectra show low-loss guidance from across much of the ultraviolet range from 190 - 400 nm, except for a high loss resonance band at 245 -- 265 nm. The minimum attenuation was 0.13 dB/m at 235 nm and 0.16 dB/m at 325 nm. Similar reported fibres had attenuations of 0.08 dB/m at 214nm \cite{YuUV} and 0.050, 0.0097 dB/m at 290, 369 nm \cite{osorio2023}. The pattern of bands indicates that 255 nm is the \break \textit{m} = 2 resonance. Eq. \ref{eqn1} then gives a wall thickness of \(\sim\) 225 nm. (The resolution of our SEM images was insufficient for us to measure the wall thickness directly with any confidence.)

\begin{figure}[h]
\centering
\includegraphics[width=1\linewidth]{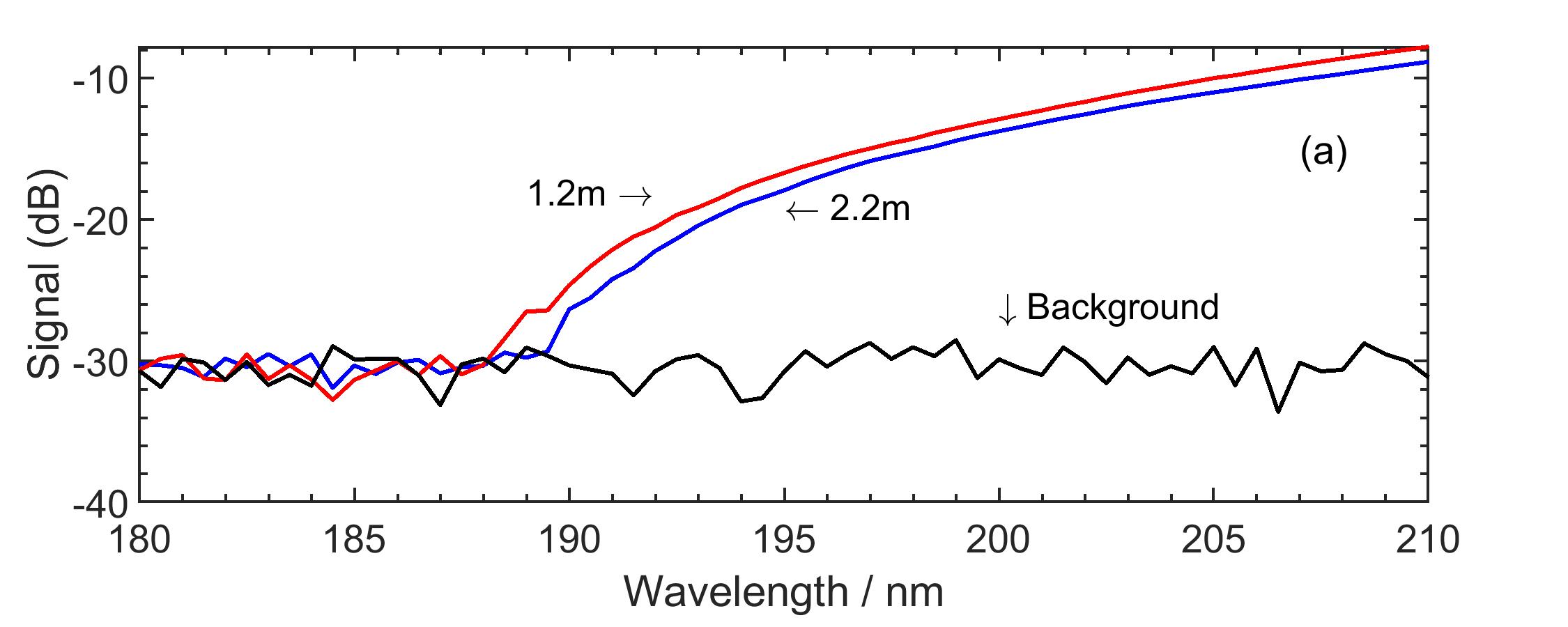}
\centering
\includegraphics[width=1\linewidth]{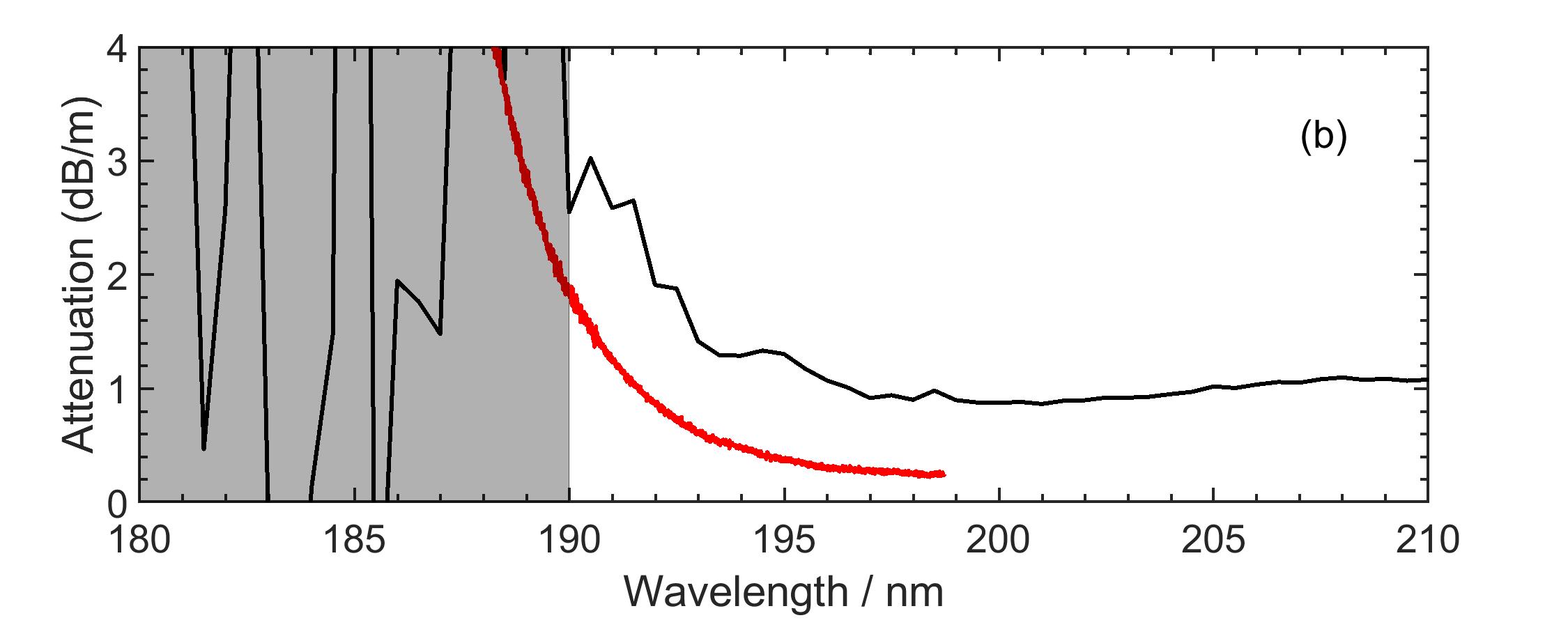}
\caption{(a) Transmission spectra for 2.2 m and 1.2 m of fibre with a background scan obtained by blocking the light source. (b) The black curve is attenuation calculated from the transmission spectra. Shaded regions indicate no signal above noise. The red curve is the estimated absorption due to water vapour in air for 50\% relative humidity at 20\(^{\circ}\)C using absorption data taken from \cite{waterabs1997,waterabs2}.}
\label{fig:shortcutback}
\end{figure}

A shorter cutback from 2.2 m to 1.2 m was performed to examine wavelengths below 200 nm, where loss is higher, Fig.\ref{fig:shortcutback}. Light is guided at wavelengths down to 190 nm, beyond which absorption by oxygen and water vapour \cite{oxabs1953,waterabs1997} prevents measurements in air. The higher attenuations compared to Fig. \ref{fig:longcutback} are consistent with the persistence of higher-order modes in the shorter fibre, along with absorption by water vapour (an estimate of which is plotted with the measured attenuation in Fig. \ref{fig:shortcutback} for room conditions close to our experiment). Based on the wall thickness inferred above and including material dispersion, Eq.\ref{eqn1} indicates the next high loss resonance should be centered on \(\sim\)185 nm.

\FloatBarrier
\section{Bend Loss}

\begin{figure}[h]
\centering
\includegraphics[width=1\linewidth]{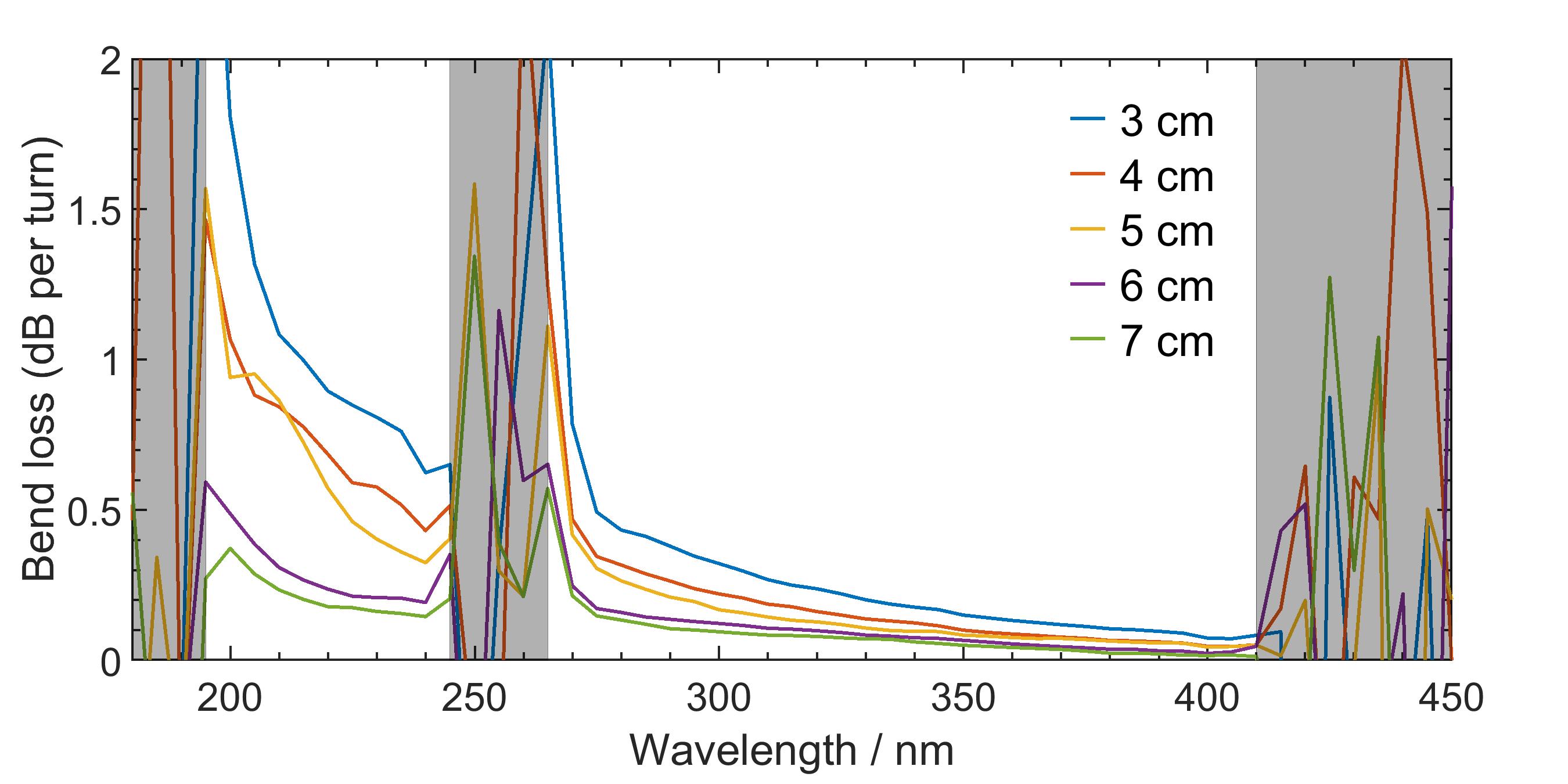}
\caption{Bend loss spectra for 5 turns around discs of the indicated radii. Shaded regions indicate no signal above noise.}
\label{fig:bendloss}
\end{figure}

Many applications require the delivery of light through tight bends with low loss. Using the rough guide that the core should be 20 – 30 times the guided wavelength, the range 600 – 400 nm would be ideally guided by this fibre with its \(\sim\)12 µm core, with increasing bend loss expected at shorter wavelengths.\newline
\indent The spectral transmission through \(\sim\)4 m of loosely wound fibre in \(\sim\)45 cm diameter loops was measured before and after some was wound through 5 turns around a smaller disc. To check that input coupling had not changed, the fibre was returned to the original loose configuration and a final spectrum recorded. Good agreement between the first and last scans indicated the calculated loss was due to the difference in bend conditions only. Bend loss is plotted in Fig. \ref{fig:bendloss} for radii from 3 – 7 cm.

\FloatBarrier
As expected, the shorter wavelengths exhibit higher bend loss than the longer wavelengths. Compared to another recent UV guiding fibre \cite{Leroi2023}, this fibre displays greatly improved bend resistance, with 85\% and 95\% transmission at 235 and 325 nm for a full turn of 3 cm radius.

\section{Discussion}
The fabrication challenges associated with UV guiding fibres usually require that the UV bandwidth is sacrificed by using higher-order bands and/or a larger core than optimal for bend loss. These compromises reduce the overall utility of the fabricated fibres and limit their applications. In this work we have shown that the fibre fabrication process can be controlled sufficiently for these compromises to be unnecessary, yielding long lengths of fibre guiding with low loss across most of the UV spectrum down to the edge of the vacuum UV while limiting bend loss. This was achieved using a single intermediate preform stage through high drawing stresses enabled by a faster and hotter draw, permitting higher potential yields than the more-complicated two-stage process reported recently \cite{Davidson23}. With elimination of water vapour in the fibre core we expect transmission to extend into the vacuum UV via higher order bands. 

With guidance across nearly the entire UV range in air of 190 – 400 nm, this fibre makes the fibre-optic delivery of UV light practical. Commonly frequency-converted laser wavelengths such as 355, 343, 214 and 206 nm all lie within the guidance windows of this fibre, enabling laser machining, ultrafast and Raman spectroscopy \cite{Kotsina2019,wolynski2011,Shutov19} with a variety of sources while avoiding structural damage \cite{Lekosiotis2023} or solarisation \cite{YuUV}. The guidance of 190 - 200 nm light in a non-tapered fibre for the first time further expands the range of AR-HCFs, with VUV guidance expected in higher order bands in the absence of oxygen or water vapour.

Further work within evacuated or purged fibres will be needed to ascertain the fundamental limits of AR-HCFs in the vacuum UV. In addition, we expect improvements to the fibre structure through the optimization of resonator size to further reduce the attenuation.

\section{Backmatter}
\begin{backmatter}
\bmsection{Funding}This work was funded by the EPSRC under grant EP/T020903/1.

\bmsection{Disclosures} The authors declare no conflicts of interest.

\bmsection{Data Availability Statement} The relevant data is available from the \href{https://doi.org/10.15125/BATH-01333}{University of Bath Research Data Archive}
\end{backmatter}

\bibliography{sample}

\end{document}